\documentclass[prl,twocolumn,showpacs,preprintnumbers,amsmath,amssymb,a4paper]{revtex4}
\usepackage{epsf}
\usepackage{amsfonts}
\usepackage{amssymb}
\usepackage{amsthm}
\usepackage{graphicx}
\def\be{\begin{equation}}
\def\ee{\end{equation}}
\def\bea{\begin{eqnarray}}
\def\eea{\end{eqnarray}}

\def\ve{\varepsilon}
\def\atanh{\mathop{{\rm atanh}}}
\def\etal{{\it et al.}\ }
\begin{document}

%\date{\today}
\title{The Lipkin-Meshkov-Glick model:\ `quasi-local' quantum criticality in nuclear physics}
\author{C. A. Hooley$^1$ and P. D. Stevenson$^2$}
\affiliation{$\mbox{}^1$Scottish Universities Physics Alliance, School of Physics and Astronomy, University of St Andrews, North Haugh, St Andrews, Fife KY16 9SS, United Kingdom}
\affiliation{$\mbox{}^2$Department of Physics, University of Surrey, Guildford GU2 7XH, United Kingdom}

\begin{abstract}
Motivated by recent work on local quantum criticality in condensed matter systems, we study the Lipkin-Meshkov-Glick (LMG) model of nuclear physics as a simple model of a kind of `quasi-local' quantum criticality.  We identify a new crossover temperature, $T^{*}(V,W)$, between linear and nonlinear dynamics, which is analogous to the crossover between the renormalized classical and quantum critical regimes in the condensed-matter case.  This temperature $T^{*}$ typically vanishes logarithmically as the quantum phase transition is approached, except near the quantum tricritical point where it becomes linear.  We also note a further analogy with condensed-matter quantum criticality:\ the LMG model exhibits quantum order-by-disorder phenomena, of the type often associated with phase reconstruction near quantum critical points.
\end{abstract}

%\keywords{Two-channel Kondo effect, Luttinger liquid, Coulomb blockade}

\maketitle

{\it Introduction.}
Theories of `local quantum criticality' have been current in the condensed matter community for over a decade.  They were initially inspired by neutron-scattering measurements on CeCu$_{6-x}$Au$_x$ in 1998 \cite{Schroeder1998}, which showed soft modes occupying a significant region of the Brillouin zone, rather than the small patch predicted by conventional theories of metallic quantum criticality \cite{HM}.  Theories of such local quantum criticality soon followed, with early examples due to Si \etal \cite{Si1999} and Coleman \cite{Coleman1999}.

Another theme that has emerged over the past decade and a half is the significance of tricriticality and phase reconstruction, particularly near ferromagnetic quantum critical points.  The observation that the momentum-dependence of the magnetic susceptibility of Fermi liquids contains non-analytic terms was made as early as 1977 by Geldart and Rasolt \cite{Geldart1977}.  This was rediscovered in 1997 by Belitz, Kirkpatrick, and Vojta, who discussed its implications for the low-temperature behavior of metallic ferromagnets \cite{Belitz1997}.  The emerging picture, now supported by a large body of experimental work \cite{Pfleiderer2004,Huxleyetal2007}, is that a second-order transition to metallic ferromagnetism generically develops a tricritical point at non-zero temperature, below which (at zero applied magnetic field) the transition becomes first-order.  This feature, however, may be occluded or supplemented by phase reconstruction in the vicinity of the quantum critical point \cite{Karahasanovic2012}.

It is always desirable to have toy models that exhibit phenomena analogous to those in more complex condensed-matter contexts.  In this Letter, we shall show that the Lipkin-Meshkov-Glick (LMG) model --- originally a model of monopole oscillations in the $^{16}$O nucleus --- shows several of the abovementioned features.  It may thus prove a fruitful ground for controlled analytic and numerical study of `quasi-local quantum criticality', a term we define below.

The LMG model has long been studied in the nuclear physics community.  It was introduced in 1959 by Fallieros \cite{Fallieros1959}, and subsequently studied by Volkov \cite{Volkov1963}, before springing to prominence with the work of Lipkin, Meshkov, and Glick in 1965 \cite{LMG1,LMG2}, who considered it as a non-trivial correlated model against which various approximation schemes could be tested.

{\it Hamiltonian, and review of known results.}  The Hamiltonian of the LMG model is:
\bea
H & = & \frac{\varepsilon}{2} \sum_{p\sigma} \sigma a^\dagger_{p\sigma} a_{p\sigma}
+ \frac{\tilde W}{2} \sum_{pq\sigma} a^\dagger_{p\sigma} a^\dagger_{q{\bar\sigma}} a_{q\sigma} a_{p{\bar\sigma}} \nonumber \\
& & \qquad \qquad \qquad \qquad + \frac{\tilde V}{2} \sum_{pq\sigma} a^\dagger_{p\sigma} a^\dagger_{q\sigma} a_{q{\bar \sigma}} a_{p{\bar \sigma}},
\label{lmgham1}
\eea
where the $a_{p\sigma}$ are fermionic annihilation operators, $\sigma=\pm 1$ is a spin-like index denoting the nuclear shell in which the fermion is, ${\bar \sigma}$ represents the opposite spin to $\sigma$, and $p=1,2,\ldots,N_p$ is an auxiliary quantum number distinguishing between a large number of degenerate levels within each shell.  In the condensed-matter context, this would represent the number of degrees of freedom across which the physical response of the system near the quantum critical point is coherent, and could presumably be written $(\xi/a)^d$, where $\xi$ is a sort of coherence length, $a$ the crystal lattice spacing, and $d$ the dimensionality of the lattice.  It is in this sense that we call the quantum criticality studied here `quasi-local':\ the scale $\xi$ is large compared to the lattice spacing, but nonetheless does not diverge as the quantum critical point is approached.

The ${\tilde W}$ interaction in (\ref{lmgham1}) represents an exchange of particles between the lower- and higher-energy shells, while ${\tilde V}$ represents pair-tunnelling between shells (a sort of Josephson term).  They may be thought of as approximate representations of the interactions between electrons in the patch.  The Hamiltonian clearly conserves the number of particles, $N \equiv \sum_{p\sigma} a^\dagger_{p\sigma} a_{p\sigma}$.  It is sometimes stipulated that $N=N_p$ (the `half-filled' case), but we shall consider all possible values of $N$ (i.e.\ all possible electron densities within the coherent patch).

The literature on the LMG model is extensive \cite{Cejnar10}.  The zero-temperature phase diagram has been obtained in the thermodynamic limit \cite{Dusuel2005,Ribeiro2007} and finite-size corrections analysed \cite{Dusuel2004,Dusuel2005}.  Non-zero-temperature properties of the model have also been studied \cite{Scherer2009}, as have properties of the zero-temperature entanglement entropy \cite{vidal2007} and the negativity \cite{wichterich2010}.  Frequently these studies confine themselves to a particular line or region in the two-parameter space $( {\tilde V},{\tilde W} )$ \cite{WSG2005}, but some works consider the whole plane.  It should also be noted that the ${\tilde V}=0$ line of the model is equivalent to the Dicke model \cite{dicke1954}, about which much is known \cite{vidal2006}.  The linear-to-nonlinear crossover we shall find below therefore also constitutes a new crossover scale in the Dicke model phase diagram.

There has, to our knowledge, not yet been a study of the non-zero-temperature properties of the model over the whole $({\tilde V},{\tilde W})$-plane in the case where $N_p$ is large but finite.  In this Letter we carry out that study, with particular emphasis on the nature of the crossover from linear to nonlinear dynamics that occurs as the quantum phase transitions in the $({\tilde V},{\tilde W})$-plane are approached at non-zero temperature.

The large-$N_p$ limit must be taken with care.  Since the interaction terms in (\ref{lmgham1}) cause every level to interact equally with every other (producing a result $\sim N_p^2$ provided that $N/N_p$ is finite), we must compensate by sending ${\tilde V}$ and ${\tilde W}$ to zero in the following way:
\be
{\tilde V} = \frac{V}{N_p}, \quad {\tilde W} = \frac{W}{N_p},
\ee
with $V$ and $W$ held constant as $N_p \to \infty$.  Then the energy remains proportional to $N_p$, with sub-dominant corrections $O(1)$.  However, as we shall see below, these sub-dominant pieces are crucial in breaking a degeneracy in $N$ in certain regions of the ground-state phase diagram.

A key observation, made in one of the original papers \cite{LMG1}, is that the Hamiltonian (\ref{lmgham1}) may be rewritten in terms of the following pseudospin operators:
\be
J_z \equiv \frac{1}{2} \sum_{p\sigma} \sigma a^\dagger_{p\sigma} a_{p\sigma}, \quad
J_{+} \equiv \sum_{p} a^\dagger_{p1} a_{p{\bar 1}}, \quad
J_{-} = J_{+}^\dagger.
\ee
These obey the standard angular momentum commutation relations.  (Note that we have adopted units in which $\hbar = 1$.)  The pseudospin version of the Hamiltonian may easily be shown to be:
\be
H = \varepsilon J_z + \frac{V}{2N_p} \left( J_{+}^2 + J_{-}^2 \right)
+ \frac{W}{2N_p} \left( J_{+} J_{-} + J_{-} J_{+} - N \right), \label{lmgquantum}
\ee
where $N$ is the particle-number operator defined above.  This form of the Hamiltonian is most useful for direct quantum treatments; for classical and semi-classical approaches it is preferable to rewrite it again using the definitions $J_{\pm} = J_x \pm iJ_y$:
\be
H = \varepsilon J_z + \frac{V}{N_p} \left( J_x^2 - J_y^2 \right)
+ \frac{W}{N_p} \left( J_x^2 + J_y^2 - \frac{N}{2} \right).
\label{lmgclassical}
\ee
As well as the particle number, $N$, this Hamiltonian also clearly conserves the magnitude of the pseudospin:
\be
{\bf J}^2 \equiv J_x^2 + J_y^2 + J_z^2 = J(J+1),
\ee
where $J$ is an integer between $0$ and $N/2$.  (We assume here and henceforth that $N$ and $N_p$ are even.)

\begin{figure}
\begin{center}
\includegraphics[width=8cm]{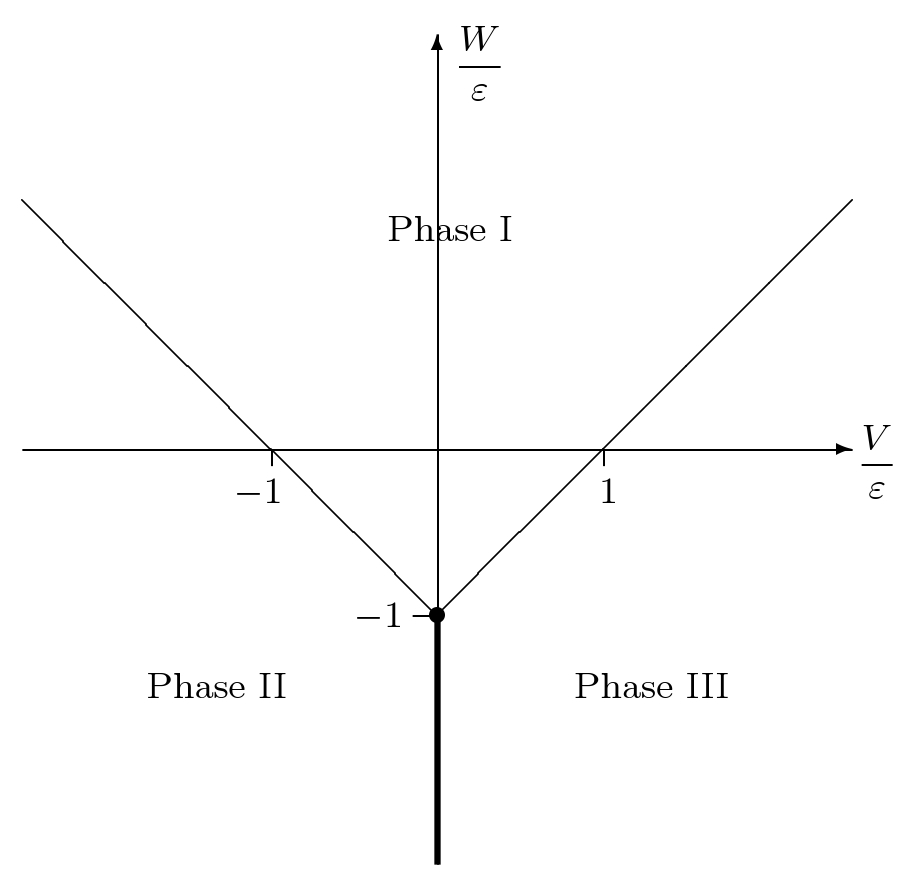}
\end{center}
\caption{The ground state phase diagram of the LMG model in the classical ($N,J \to \infty$) limit.  The bold transition line at $V=0$ is first-order, while the other two are second-order; they meet at a tricritical point when $(V,W)=(0,-\ve)$.}
\label{gsphasediag}
\end{figure}

{\it Classical phase diagram.}  Firstly we shall summarize the analysis of the Hamiltonian (\ref{lmgclassical}) in the limit $J,N \to \infty$ \cite{Dusuel2005}.  Note that these limits, while not independent, are nonetheless not the same: $J \to \infty$ implies $N \to \infty$, but not vice versa.  In the $J \to \infty$ limit, the operators may be replaced by classical vectors, and the energy minimized straightforwardly.  We choose the following parameterization:
\bea
J & = & \frac{J_{\rm max}}{2} \left( 1-\cos\alpha \right), \\
(J_x,J_y,J_z) & = & \left( J\sin\theta \cos\phi,J\sin\theta \sin\phi,J\cos\theta \right).
\eea
The maximum possible value of $J$, denoted $J_{\rm max}$, is of course a function of $N$:
\be
J_{\rm max} = \left\{ \begin{array}{lll}
\frac{N}{2} & \quad & 0 \leqslant N \leqslant N_p, \\
& & \\
N_p - \frac{N}{2} & & N_p < N \leqslant 2N_p. \end{array} \right.
\ee
Minimizing the energy simultaneously with respect to $\alpha$, $\theta$, and $\phi$, one obtains the ground state phase diagram shown in Fig.~\ref{gsphasediag}.
The three phases shown in the diagram are characterized by the following behaviors of the pseudospin ${\bf J}$:

Phase I.  Full spin, oriented in the negative $z$ direction.  This corresponds to the parameter values $\alpha=\pi$, $\theta=\pi$, $\phi$ undetermined.  This ground state is non-degenerate.

Phase II.  Full spin, canting from the negative $z$ direction (near the transition line) to either the negative or positive $x$ direction (deep in the phase).  This corresponds to the parameter values $\alpha=\pi$, $\phi=0$ or $\pi$, and
\be
\theta = \arccos \left( \frac{\ve}{W+V} \right).
\ee
This shows in particular that the transition is second-order:\ there is no jump in the spin's angle of orientation as the boundary between phases I and II is crossed.  In phase II, the ground state is always doubly degenerate.

Phase III.  As phase II, but with $\phi=\pm \pi/2$.

The phase boundary between phases II and III is first-order, since while $\alpha$ and $\theta$ are continuous across it, it involves a discontinuous jump of the parameter $\phi$, corresponding to a reorientation of the spin from the $x$- to the $y$-axis.  However, it is a peculiar sort of first-order transition, since at the transition all values of $\phi$ become degenerate, and hence despite being first-order it does have associated soft modes.  This emergent U(1) symmetry at $V=0$ is nothing but the phase of the coherent photon field in the superradiant phase of the Dicke model, with the quantum tricritical point at $(V,W)=(0,-\ve)$ corresponding to the superradiance transition.

{\it Crossover to non-linear dynamics.} We now proceed to analyse the finite-$N_p$ model at non-zero temperature.  It is natural to choose a Holstein-Primakoff representation \cite{HolsteinPrimakoff1940} of the pseudospin, which we define with respect to phase I, i.e.\ with reference to a full spin oriented in the negative $z$-direction:
\be
J_z \equiv -J + b^\dagger b, \quad
J_+ \approx \sqrt{2J} \,b^\dagger, \quad J_- \approx \sqrt{2J} \,b,
\ee
where the boson operators $b$ and $b^\dagger$ obey the usual commutation relations $[b,b^\dagger]=1$, and the linear approximation has been made.  Although $J$ is formally variable, we shall here treat it as fixed at $N_p/2$.

Substituting this approximation into (\ref{lmgquantum}), and applying the commutation relations for the $b$-operators, we obtain that
\be
H = -\frac{\varepsilon N_p}{2} + \left( \varepsilon+W \right) b^\dagger b + \frac{V}{2} \left( (b^\dagger)^2 + (b)^2 \right). \label{bosonham}
\ee
It is clear from (\ref{bosonham}) that the tricritical point $(V,W)=(0,-\varepsilon)$ corresponds to the point where the boson energy becomes negative, signalling an instability which mathematically invalidates the linear approximation, and physically corresponds to the superradiance transition.  To extend the analysis to non-zero $V$, we must make a Bogolyubov rotation \cite{Bogolyubov1947} to eliminate the anomalous terms $b^2$ and $(b^\dagger)^2$.  Such a rotation is possible only in the interval
$\left\vert \varepsilon+W \right\vert \geqslant \left\vert V \right\vert$,
i.e.\ in the area labelled `phase I' in the classical analysis above.  In this region, the result of the transformation is
\be
H = E \beta^\dagger \beta + {\rm const.},
\ee
where the boson energy $E$ is given by
\be
E = \sqrt{(\varepsilon+W)^2-V^2}.
\ee

It is a familiar feature of quantum critical theories \cite{HM,CHN} that the approach to a quantum critical point at non-zero temperature is accompanied by a crossover from `renormalized classical' to `quantum critical' behavior.  A phenomenon of the same sort takes place here:\ the dynamics of the model cross over from being approximately linear to fully nonlinear as the transition line is approached.  A simple way to obtain the location of this crossover is to ask at what temperature the linear approximation (i.e.\ the condition that $\left\langle b^\dagger b \right\rangle \ll N_p$) breaks down.  The temperature at which this happens is approximately given by
$\left\langle b^\dagger b \right\rangle_T = N_p$; from the condensed-matter point of view, this is the temperature at which the number of thermal excitations becomes equal to the size of the coherent patch.  Inserting the expressions for $b$ and $b^\dagger$ in terms of $\beta$ and $\beta^\dagger$ this becomes
\be
\frac{\ve + W}{\sqrt{(\ve+W)^2-V^2}} \left( \langle \beta^\dagger \beta \rangle_T + \frac{1}{2} \right) = N_p + \frac{1}{2}.
\ee
The value of the thermal average follows directly from the Bose-Einstein distribution, so that
\be
\frac{\ve + W}{\sqrt{(\ve+W)^2-V^2}} \left( \frac{1}{e^{\beta E} - 1} + \frac{1}{2} \right) = N_p + \frac{1}{2}
\ee
which yields the temperature
\be
T^{*} = \frac{\ve}{k_B} \frac{\sqrt{y^2-x^2}}{2} \left[
\atanh \left( \frac{\sqrt{\gamma^2-1}}{\gamma} \frac{y}{\sqrt{y^2-x^2}} \right) \right]^{-1}, \label{crittemp}
\ee
where
\be
x \equiv \frac{V}{\ve}, \quad y \equiv 1 + \frac{W}{\ve}, \quad \gamma \equiv \left( 1 - \frac{1}{(2N_p+1)^2} \right)^{-1/2}.
\ee
This expression for $T^{*}$, the crossover scale between linear and nonlinear dynamics near the quantum critical point, is the key result of this Letter.

\begin{figure}
\begin{center}
\includegraphics[width=8cm]{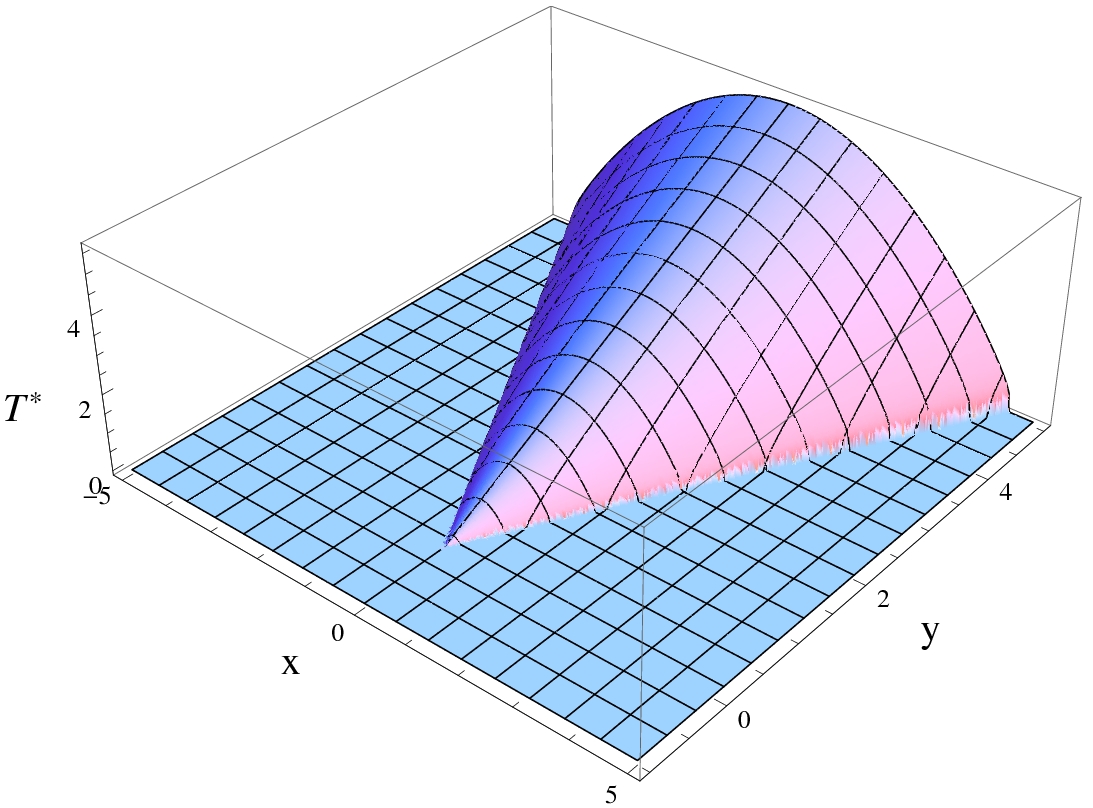}
\end{center}
\caption{The crossover temperature $T^{*}$ as a function of $x$ and $y$, the rescaled and offset interaction parameters of the LMG model.  The quantum tricritical point is at the origin; note the crossover from logarithmic to linear behavior of $T^{*}$ as this point is approached.  The line $x=0$ corresponds to the Dicke model.}
\label{tstar}
\end{figure}

The first thing to observe about (\ref{crittemp}) is that it vanishes not at the original phase transition $y=x$ but at $y = \gamma x$.  This is a renormalization of the position of the transition line due to quantum fluctuations, similar to those discussed in \cite{Moriya1985}.  To examine the behavior of $T^{*}$ as this renormalized transition line is approached, we set $y=\gamma x + \delta$, with $0 < \delta \ll 1$.  In this limit, provided that $x \gg \delta$, we obtain
\be
T^{*} \approx \frac{\ve}{k_B} \frac{x \sqrt{\gamma^2 - 1}}{\ln (2\gamma(\gamma^2-1)x) - \ln \delta}
\sim - \frac{1}{\ln \delta} \label{nearline}
\ee
as $\delta \to 0^{+}$.  Hence this second-order transition has an extremely narrow quantum critical cone, in contrast with the simple power laws typically observed in quantum critical theories \cite{CHN,HM}.  The expression (\ref{nearline}) becomes invalid as the tricritical point at $(x,y)=(0,0)$ is approached.  It crosses over to a much simpler behavior, which may be obtained by setting $x=0$ and then taking $0 < y \ll 1$:
\be
T^{*} \approx \frac{\ve}{k_B} \frac{y}{2} \left[ \atanh \left( \frac{\sqrt{\gamma^2-1}}{\gamma} \right) \right]^{-1} \sim y \label{neartcp}
\ee
as $y \to 0^{+}$.  Thus, perhaps surprisingly, power-law behavior is recovered at the tricritical point $y=x=0$ despite being absent along the rest of the critical line $y = \gamma x$.  The full behavior of (\ref{crittemp}) is shown in Fig.~\ref{tstar}.

{\it Phase reconstruction and quantum order-by-disorder.}  The phase diagram of the LMG model also contains another phenomenon reminiscent of condensed-matter quantum criticality:\ phase reconstruction of the quantum order-by-disorder type \cite{BAsol}.  To expose it, let us slightly rewrite the Hamiltonian (\ref{lmgclassical}) with $V=0$ as
\be
H = \varepsilon J_z + \frac{W}{N_p} \left( {\bf J}^2 - J_z^2 - \frac{N}{2} \right).
\ee
Classically, it is possible to make ${\bf J}^2 - J_z^2$ zero by orienting the spin along the positive or negative $z$-direction; quantum mechanically, however, this zero is achieved only in the singlet state, where ${\bf J}^2=0$.

This effect manifests itself at large positive values of $W/\varepsilon$, where to a first approximation one may neglect the $\varepsilon$ term entirely.  The Hamiltonian is then minimized by (a) choosing a singlet state for the pseudospin, and (b) manufacturing this singlet state from the largest possible number of particles; in this case, that number is $N=2N_p$, corresponding to full occupation of all the levels in the original LMG model.  The energy of this singlet state is therefore simply
$E_{\rm singlet} = -W$.
By comparison, the energy of the phase I state at $V=0$ is
$E_{\rm full{-}spin} = -\varepsilon N_p/2$.
Hence the transition from phase I to the singlet phase occurs when $W/\varepsilon = N_p/2$.  Further analysis shows that there are no intervening phases; the transition occurs directly from full- to zero-spin, via a rather interesting quantum critical point.

For non-zero $V$, quantum fluctuations shift the critical value of $W$; to leading non-zero order in perturbation theory, the resulting behavior is given by:
\be
W_c = \frac{\varepsilon N_p}{2} + \frac{V^2}{\varepsilon N_p} \left( 1 - \frac{1}{N_p} \right).
\ee
The fact that $W_c$ increases with increasing $V$ represents a pseudo-entropic favouring of the full-spin state, of the type recently discussed by Conduit {\it et al.} \cite{ConduitGreenSimons2009}.

{\it Summary.}
In this Letter, we have pointed out that three features associated with condensed-matter quantum criticality are also present in the `quasi-local' version of the Lipkin-Meshkov-Glick model.  The first is tricriticality --- see Fig.~\ref{gsphasediag}.  The second is a crossover between renormalized classical and quantum critical behavior, or, in the language of dynamical systems, between linear and nonlinear dynamics.  The third is the presence of quantum order-by-disorder effects in the low-temperature phase diagram.  Our use of a `finite-size but infinite-range' model to mimic the effect of a large but non-divergent spatial correlation length, as seen in CeCu$_{6-x}$Au$_x$, allows us to obtain a full analytic form for the crossover temperature $T^{*}$.

We thank Professor A. J. Schofield and Dr J. M. J. Keeling for helpful discussions, and gratefully acknowledge financial support from the EPSRC (UK), the STFC (UK), the Nuffield Foundation and the Scottish Universities Physics Alliance.

\end{document}